\documentstyle[12pt,epsfig]{article}

\def\be{\begin{equation}}
\def\ee{\end{equation}}

\def\IP{\hbox{\rm I\kern -1.6pt{\rm P}}}
\def\IC{{\hbox{\rm C\kern-.58em{\raise.53ex\hbox{$\scriptscriptstyle|$}}
    \kern-.55em{\raise.53ex\hbox{$\scriptscriptstyle|$}} }}}
\def\IN{\hbox{I\kern-.2em\hbox{N}}}
\def\IR{\hbox{\rm I\kern-.2em\hbox{\rm R}}}
\def\ZZ{\hbox{{\rm Z}\kern-.3em{\rm Z}}}
\def\IT{\hbox{\rm T\kern-.38em{\raise.415ex\hbox{$\scriptstyle|$}} }}

\newtheorem{theorem}{Theorem}[section]
\newtheorem{lemma}[theorem]{Lemma}

\newtheorem{corollary}[theorem]{Corollary}

\begin{document}

\title{Scaling Dynamics of a Massive Piston in a Cube
Filled With Ideal Gas: Exact Results}
\author{N. Chernov$^{1,4}$,
J.~L.~Lebowitz$^{2,4}$, and Ya.~Sinai$^{3}$}
\date{\today}
\maketitle

\begin{center}
{\it Dedicated to Robert Dorfman on the occasion of his 65$^{\,\rm
th}$ birthday}
\end{center}

\begin{abstract}
We continue the study of the time evolution of a system consisting
of a piston in a cubical container of large size $L$ filled with
an ideal gas. The piston has mass $M\sim L^2$ and undergoes
elastic collisions with $N\sim L^3$ gas particles of mass $m$. In
a previous paper, Lebowitz, Piasecki and Sinai \cite{LPS}
considered a scaling regime, with time and space scaled by $L$, in
which they argued heuristically that the motion of the piston and
the one particle distribution of the gas satisfy autonomous
coupled differential equations. Here we state exact results and
sketch proofs for this behavior.

\footnotetext[1]{Department of Mathematics,
University of Alabama at Birmingham, Alabama 35294}
\footnotetext[2]{Department of Mathematics, Rutgers
University, New Jersey 08854}
\footnotetext[3]{Department of Mathematics, Princeton University,
New Jersey 08544}
\footnotetext[4]{Current address: Institute for Advanced Study,
Princeton, NJ 08540}

\end{abstract}

\renewcommand{\theequation}{\arabic{section}.\arabic{equation}}

\section{The model and main results}
\label{secI} \setcounter{equation}{0}

This paper is a continuation of \cite{LPS}, where deterministic
scaled equations describing the dynamics of a massive piston
in a cubical container filled with ideal gas were given. Here
we state exact conditions on the validity of those equations and
outline the arguments. Full proofs will be published in a separate
paper \cite{CLS}.

We refer the reader to \cite{LPS} and to \cite{CLS} as well as
\cite{G,GF,GP,KBM,Li} for a detailed description of the problem of
a massive piston moving in a cylinder. Here we just recall
necessary facts.

Consider a cubical domain $\Lambda_L$ of size $L$ separated into
two parts by a wall (piston), which can move freely without
friction inside $\Lambda_L$.  Each part of $\Lambda_L$ is filled
by a noninteracting gas of particles, each of mass $m$.  The
piston has mass $M=M_L$ and moves along the $x$-axis under the
action of elastic collisions with the particles. The size $L$ of
the cube is a large parameter of the model, and we are interested
in the limit behavior as $L\to\infty$. The mass $m$ of gas
particles is fixed. We will assume that $M$ grows proportionally
to $L^2$ and the number of gas particles $N$ is proportional to
$L^3$, while their velocities remain of order one.

The position of the piston at time $t$ is specified by a single
coordinate $X=X_L(t)$, $0 \leq X \leq L$, its velocity is then
given by $V=V_L(t)=\dot{X}_L(t)$.  Since the components of the
particle velocities perpendicular to the $x$-axis play no role in
the dynamics, we may assume that each particle has only one
coordinate, $x$, and one component of velocity, $v$, directed
along the $x$-axis.

When a particle with velocity $v$ hits the piston with velocity
$V$, their velocities after the collision, $v^\prime$ and
$V^\prime$, respectively, are given by
\be
    V^\prime = (1-\varepsilon)V + \varepsilon v
      \label{V'}
\ee
\be
    v^\prime = -(1-\varepsilon)v + (2-\varepsilon)V
      \label{v'}
\ee
where $ \varepsilon = 2m/(M+m)$. We assume that
$M+m=2mL^2/a$, where $a>0$ is a constant, so that
\be
    \varepsilon = \frac{2m}{M+m}=\frac{a}{L^2}
      \label{varepsMm}
\ee
When a particle collides with a wall at $x=0$ or $x=L$, its
velocity just changes sign.

The evolution of the system is completely deterministic, but one
needs to specify initial conditions. We shall assume that the
piston starts at the midpoint $X_L(0)=L/2$ with zero velocity
$V_L(0)=0$. The initial configuration of gas particles is chosen
at random as a realization of a (two-dimensional) Poisson process
on the $(x,v)$-plane (restricted to $0\leq x\leq L$) with density
$L^2p_L(x,v)$, where $p_L(x,v)$ is a function satisfying certain
conditions, see below, and the factor of $L^2$ is the
cross-sectional area of the container. In other words, for any
domain $D\subset [0,L]\times\IR^1$ the number of gas particles
$(x,v)\in D$ at time $t=0$ has a Poisson distribution with
parameter $\lambda_D=L^2\int\!\int_Dp_L(x,v)\, dx\, dv$.

Let $\Omega_L$ denote the space of all possible configurations of
gas particles in $\Lambda_L$. For each realization
$\omega\in\Omega_L$ the piston trajectory will be denoted by
$X_L(t,\omega)$ and its velocity by $V_L(t,\omega)$.

As $L\to\infty$, space and time are rescaled as
\be
     y=x/L\ \ \ \ {\rm and}\ \ \ \  \tau=t/L.
     \label{ytau}
\ee
which is a typical rescaling for the hydrodynamic limit transition
(see \cite{LPS,CLS} for motivation and physical discussion). We
call $y$ and $\tau$ the {\em macroscopic} (``slow'') variable, as
opposed to the original {\em microscopic} (``fast'') $x$ and $t$.
Now let
\be
     Y_L(\tau,\omega)=X_L(\tau L,\omega)/L,
       \ \ \ \ \ \ \ \ \ \
       W_L(\tau,\omega)=V_L(\tau L,\omega)
          \label{YWL}
\ee
be the position and velocity of the piston in the macroscopic
variables.
%The initial conditions are $Y_L(0)=0.5$ and $W(0)=V(0)$.
The initial density $p_L(x,v)$ satisfies
$$
      p_L(x,v) = \pi_0(x/L,v)
$$
where the function $\pi_0(y,v)$ is independent of $L$. Without
loss of generality, assume that $\pi_0$ is normalized so that
$$
     \int_0^1\int_{-\infty}^{\infty}\pi_0(y,v)\, dv\, dy = 1
$$
Then the mean number of particles in the entire container
$\Lambda_L$ is equal exactly to
$$
       E(N)=\int_0^L\int_{-\infty}^{\infty} L^2p_L(x,v)\,dv\, dx = L^3
$$

In order to describe the dynamics by differential equations, we
assume that the function $\pi_0(y,v)$ satisfies several technical
requirements stated below.

\begin{itemize}
\item[(P1)] {\em Smoothness}. $\pi_0(y,v)$ is a piecewise $C^1$
function with uniformly bounded partial derivatives, i.e.\
$|\partial\pi_0/\partial y|\leq D_1$ and $|\partial\pi_0/\partial
v|\leq D_1$ for some $D_1>0$.

\item[(P2)] {\em Discontinuity lines}. $\pi_0(y,v)$ may be
discontinuous on the line $y=Y_L(0)$ (i.e., ``on the piston''). In
addition, it may have a finite number ($\leq K_1$) of other
discontinuity lines in the $(y,v)$-plane with strictly positive
slopes (each line is given by an equation $v=f(y)$ where $f(y)$ is
$C^1$ and $0<c_1<f'(y)<c_2<\infty$).

\item[(P3)] {\em Density bounds}. Let
\be
         \pi_0(y,v)>\pi_{\min}>0\ \ \ \ \ \ {\rm for}\ \ v_1<|v|<v_2
                \label{pmin0}
\ee
for some $0<v_1<v_2<\infty$, and
\be
   \sup_{y,v}\pi_0(y,v)=\pi_{\max}<\infty
      \label{pmax0}
\ee
The requirements (\ref{pmin0}) and (\ref{pmax0}) basically mean
that $\pi_0(y,v)$ takes values of order one.

\item[(P4)] {\em Velocity ``cutoff''}. Let
\be
     \pi_0(y,v) = 0, \quad {\rm if} \quad |v| \leq v_{\rm min}
     \quad {\rm or} \quad |v| \geq v_{\max}
        \label{0cutoff}
\ee
with some $0<v_{\min}<v_{\max}<\infty$. This means that the speed
of gas particles is bounded from above by $v_{\max}$ and from
below by $v_{\min}$.

\item[(P5)] {\em Approximate pressure balance}. $\pi_0(y,v)$ must
be nearly symmetric about the piston, i.e.
\be
         |\pi_0(y,v)-\pi_0(1-y,-v)|<\varepsilon_0
             \label{0symmetry}
\ee
for all $0<y<1$ and some sufficiently small $\varepsilon_0>0$.
\end{itemize}

The requirements (P4) and (P5) are made to ensure that the piston
velocity $|V_L(t,\omega)|$ will be smaller than the minimum speed
of the particles, with probability close to one, for times
$t=O(L)$. Such assumptions were first made in \cite{LPS}.

We think of $D_1$, $K_1$, $c_1$, $c_2$, $v_1$, $v_2$, $v_{\min}$,
$v_{\max}$, $\pi_{\min}$ and $\pi_{\max}$ in (P1)--(P4) as fixed
(global) constants and $\varepsilon_0$ in (P5) as an adjustable
small parameter. We will assume throughout the paper that
$\varepsilon_0$ is small enough, meaning that
$$
       \varepsilon_0<\bar{\varepsilon}_0
         (D_1,K_1,c_1,c_2,v_1,v_2,v_{\min},v_{\max},\pi_{\min},\pi_{\max})
$$
It is important to note that the hydrodynamic limit does {\em not}
require that $\varepsilon_0\to 0$. The parameter $\varepsilon_0$
stays positive and fixed as $L\to\infty$.

Here is our main result:

\begin{theorem}
There is an $L$-independent function $Y(\tau)$ defined for
all $\tau\geq 0$ and a positive
$\tau_{\ast}\approx 2/v_{\max}$ (actually, $\tau_{\ast}\to
2/v_{\max}$ as $\varepsilon_0\to 0$), such that
\be
   \sup_{0\leq\tau\leq\tau_{\ast}}
        |Y_L(\tau,\omega) - Y(\tau)| \to 0
        \label{YY}
\ee
and
\be
   \sup_{0\leq\tau\leq\tau_{\ast}}
        |W_L(\tau,\omega) - W(\tau)| \to 0
       \label{WW}
\ee
in probability, as $L\to\infty$. Here $W(\tau)=\dot{Y}(\tau)$.
\label{tmmain}
\end{theorem}

This theorem establishes the convergence in probability of the
random functions $Y_L(\tau,\omega),W(\tau,\omega)$ characterizing
the mechanical evolution of the piston to the deterministic
functions $Y(\tau),W(\tau)$, in the hydrodynamic limit
$L\to\infty$. \medskip

The functions $Y(\tau)$ and $W(\tau)$ satisfy certain
(Euler-type) differential equations stated in the next section.
Those equations have solutions for all $\tau\geq 0$, but we can only
guarantee the convergence (\ref{YY}) and (\ref{WW}) for
$\tau<\tau_{\ast}$. What happens for $\tau>\tau_{\ast}$,
especially as $\tau\to\infty$, remains an open problem. Some
experimental results and heuristic observations in this direction
are presented in \cite{CL} and discussed in Section~4.
\medskip

\noindent{\bf Remark 1}. The function $Y(\tau)$ is at least $C^1$
and, furthermore, piecewise $C^2$ on the interval
$(0,\tau_{\ast})$. Its first derivative $W=\dot{Y}$ (velocity) and
its second derivative $A=\ddot{Y}$ (acceleration) remain
$\varepsilon_0$-small: $\sup_{\tau}
|W(\tau)|\leq\,$const$\cdot\varepsilon_0$ and $\sup_{\tau}
|A(\tau)|\leq\,$const$\cdot\varepsilon_0$, see the next section.
\medskip

\noindent{\bf Remark 2}. We also estimate the speed of convergence
in (\ref{YY}) and (\ref{WW}): there is a $\tau_1>0$
($\tau_1\approx 1/v_{\max}$) such that $|Y_L(\tau,\omega)
-Y(\tau)|=O(\ln L/L)$ for $0<\tau<\tau_1$ and
$|Y_L(\tau,\omega)-Y(\tau)|=O(\ln L/L^{1/7})$ for
$\tau_1<\tau<\tau_{\ast}$. The same bounds are valid for
$|W_L(\tau,\omega)-W(\tau)|$. These estimates hold with
``overwhelming'' probability, specifically they hold for all
$\omega\in\Omega^{\ast}_L\subset\Omega_L$ such that
$P(\Omega^{\ast}_L)=1-O(L^{-\ln L})$.

\section{Hydrodynamical equations}
\label{secHE} \setcounter{equation}{0}

The equations describing the deterministic function $Y(\tau)$
involve another deterministic function -- the density of the gas
$\pi(y,v,\tau)$. Initially, $\pi(y,v,0)=\pi_0(y,v)$, and for
$\tau>0$ the density $\pi(y,v,\tau)$ evolves according to the
following rules.

\begin{itemize}
\item[(H1)] {\em Free motion}. Inside the container the density
satisfies the standard continuity equation for a noninteracting
particle system without external forces:
\be
     \left ( \frac{\partial}{\partial \tau}+
      v\, \frac{\partial}{\partial y} \right )
        \, \pi(y,v,\tau)=0
          \label{pdinside}
\ee
for all $y$ except $y=0$, $y=1$ and $y=Y(\tau)$.
\end{itemize}

\noindent Equation (\ref{pdinside}) has a simple solution
\be
      \pi(y,v,\tau)=\pi(y-vs,v,\tau-s)
         \label{pinside}
\ee
for $0<s<\tau$ such that $y-vr\notin\{0,Y(\tau-r),1\}$ for all
$r\in (0,s)$. Equation (\ref{pinside}) has one advantage over
(\ref{pdinside}): it applies to all points $(y,v)$, including
those where the function $\pi$ is not differentiable.

\begin{itemize}

\item[(H2)] {\em Collisions with the walls}. At the walls $y=0$
and $y=1$ we have
\be
       \pi(0,v,\tau) = \pi(0,-v,\tau)
          \label{pwall0}
\ee
\be
       \pi(1,v,\tau) = \pi(1,-v,\tau)
          \label{pwall1}
\ee

\item[(H3)] {\em Collisions with the piston}. At the piston
$y=Y(\tau)$ we have
\begin{eqnarray}
      \pi(Y(\tau)-0,v,\tau) &=& \pi(Y(\tau)-0,2W(\tau)-v,\tau)
      \ \ \ \ \ {\rm for}\ \ v<W(\tau)\nonumber\\
      \pi(Y(\tau)+0,v,\tau) &=& \pi(Y(\tau)+0,2W(\tau)-v,\tau)
      \ \ \ \ \ {\rm for}\ \ v>W(\tau)
        \label{ponpiston}
\end{eqnarray}
where $v$ represents the velocity after the collision and
$2W(\tau)-v$ that before the collision; here
\be
        W(\tau) = \frac{d}{d\tau}Y(\tau)
          \label{W=Y'}
\ee
is the (deterministic) velocity of the piston.

\end{itemize}

It remains to describe the evolution of $W(\tau)$. Suppose the
piston's position at time $\tau$ is $Y$ and its velocity $W$. The
piston is affected by the particles $(y,v)$ hitting it from the
right (such that $y=Y+0$ and $v<W$) and from the left (such that
$y=Y-0$ and $v>W$). Accordingly, we define the density of the
particles colliding with the piston (``density on the piston'') by
\be
    q(v,\tau;Y,W)=\left\{\begin{array}{ll}
       \pi(Y+0,v,\tau)  &  {\rm if}\ \ v<W\\
       \pi(Y-0,v,\tau)  &  {\rm if}\ \ v>W\\
          \end{array}\right .
            \label{qp}
\ee

\begin{itemize}
\item[(H4)] {\em Piston's velocity}. The velocity $W=W(\tau)$ of
the piston must satisfy the equation
\be
       \int_{-\infty}^{\infty} (v-W)^2\,{\rm sgn}
       (v-W)\, q(v,\tau;Y,W)\, dv =0
            \label{quadraticint}
\ee
\end{itemize}

We also remark that for $\tau>0$, when (\ref{ponpiston}) holds,
$$
    W(\tau) =
    \frac{\int v \pi(Y-0,v,\tau)\, dv}{\int \pi(Y-0,v,\tau)\, dv}=
    \frac{\int v \pi(Y+0,v,\tau)\, dv}{\int \pi(Y+0,v,\tau)\, dv}
$$
i.e.\ the piston's velocity is the average of the nearby particle
velocities on each side.

The system of (hydrodynamical) equations (H1)--(H4) is now closed
and, given appropriate initial conditions, should completely
determine the functions $Y(\tau)$, $W(\tau)$ and $p(y,v,\tau)$ for
$\tau>0$. To specify the initial conditions, we set
$p(y,v,0)=\pi(y,v)$ and $Y(0)=0.5$. The initial velocity $W(0)$
does not have to be specified, it comes ``for free'' as the
solution of the equation (\ref{quadraticint}) at time $\tau=0$. It
is easy to check that the initial speed $|W(0)|$ will be smaller
than $v_{\min}$, in fact $W(0)\to 0$ as $\varepsilon_0\to 0$ in
(P5). Note that if the initial conditions at $\tau=0$ do not
satisfy (\ref{pwall0})--(\ref{ponpiston}), there will be a
discontinuity in $p$ as $\tau\to 0$ (see also Remark~4 below).

Equation (\ref{quadraticint}) has a unique solution $W$ as long as
the piston interacts with some gas particles on both sides, i.e.\
as long as
$$
      \inf\{v:\, \pi(Y+0,v,\tau)>0\} \leq \sup\{v:\, \pi(Y-0,v,\tau)>0\}
$$
Indeed, the left hand side of (\ref{quadraticint}) is a continuous
and strictly monotonically decreasing function of $W$, and it
takes both positive and negative values. The solution $W(\tau)$
may not be continuous in $\tau$, though. But if $\pi(y,v,\tau)$ is
piecewise $C^1$ and has a finite number of discontinuity lines
with positive slopes (as we require of $\pi_0(y,v)$ in
Section~\ref{secI}), then $W(\tau)$ will be continuous and
piecewise differentiable.

\medskip{\bf Remark 3}. One can easily check that the total
mass ${\cal M}=\int\!\int \pi(y,v,\tau)\, dv\, dy$ and the total
kinetic energy $2E=\int\!\int v^2\pi(y,v,\tau)\, dv\, dy$ remain
constant along any solution of our system of equations (H1)--(H4).
Also, the mass in the left and right part of $\Lambda_L$
separately remains constant. Equation (\ref{quadraticint}) also
preserves the total momentum of the gas $\int\!\int v\,
\pi(y,v,\tau)\, dv\, dy$, but it changes due to collisions with
the walls.

\medskip{\bf Remark 4}. Previously, Lebowitz, Piasecki and Sinai \cite{LPS}
studied the piston dynamics under essentially the same initial
conditions as our (P1)--(P5). They argued heuristically that the
piston dynamics could be approximated by certain deterministic
equations in the original (microscopic) variables $x$ and $t$. The
deterministic equations found in \cite{LPS} correspond to our
(\ref{pinside})--(\ref{W=Y'}) with obvious transformation back to
the variables $x,t$, but our main equation (\ref{quadraticint})
has a different counterpart in the context of \cite{LPS}, which
reads
\be
    \frac{d}{dt}V(t)=a\,\left [ \int_V^{\infty} (v-V(t))^2 \pi(Y-0,v,t)\, dv
      - \int_{-\infty}^V (v-V(t))^2 \pi(Y+0,v,t)\, dv \right ]
        \label{tquadraticint}
\ee
Here $X=X(t)$ and $V=V(t)=\dot{X}(t)$ denote the deterministic
position and velocity of the piston and $\pi(x,v,t)$ the density
of the gas (the constant $a$ appeared in (\ref{varepsMm})). We
refer to \cite{LPS} for more details and a heuristic derivation of
(\ref{tquadraticint}). Since (\ref{tquadraticint}), unlike our
(\ref{quadraticint}), is a differential equation, the initial
velocity $V(0)$ has to be specified separately, and it is
customary to set $V(0)=0$. Alternatively, one can set $V(0)=W(0)$,
see \cite{CLS}. Equation (\ref{tquadraticint}) can be reduced to
(\ref{quadraticint}) in the limit $L\to\infty$ as follows. One can
show (we omit details) that (\ref{tquadraticint}) is a dissipative
equation whose solution with any (small enough) initial condition
$V(0)$ converges to the solution of (\ref{quadraticint}) during a
$t$-time interval of length $\sim\ln L$. That interval has length
$\sim L^{-1}\ln L$ on the $\tau$ axis, and so it vanishes as
$L\to\infty$, this is why we replace (\ref{tquadraticint}) with
(\ref{quadraticint}) and ignore the initial condition $V(0)$ when
working with the thermodynamic variables $\tau$ and $y$. The
equation (\ref{tquadraticint}) is not used in this paper.
\medskip

We now describe the solution of our equations (H1)-(H4) in more
detail. Assume that for some $\tau > 0$ the gas density
$\pi(y,v,\tau)$ satisfies the same requirements (P1)-(P4) as those
imposed on the initial function $\pi_0(y,v)$ in Sect.~\ref{secI},
with constants $D_1'$, $K_1'$, $c_1'$, $c_2'$, $v_1'$, $v_2'$,
$v_{\min}'$, $v_{\max}'$, $\pi_{\min}'$ and $\pi_{\max}'$, whose
values are not essential, but are independent of $\tau$.

We also assume an analogue of (P5), but this one is not so
straightforward, since the piston does not have to stay at the
middle point $y=0.5$ at any time $\tau >0$. We require that
\be
        |Y(\tau)-0.5|<\varepsilon_0'
          \label{Y05}
\ee
and for any point $(y,v)$ with $v_{\min}'\leq |v|\leq v_{\max}'$
there is another point $(y_{\ast},v_{\ast})$ ``across the piston''
(i.e. such that $(y-Y(\tau))(y_{\ast}-Y(\tau))<0$) satisfying
\be
          |y+y_{\ast}-1|<\varepsilon_0',\ \ \ \ \ \ \
          |v+v_{\ast}|<\varepsilon_0'
           \label{Rtau}
\ee
and
\be
         |\pi(y,v,\tau)-\pi(y_{\ast},v_{\ast},\tau)|<\varepsilon_0'
             \label{symmetry}
\ee
for some sufficiently small $\varepsilon_0'>0$. Actually, the map
$(y,v)\mapsto (y_{\ast},v_{\ast})$, which we denote by $R_{\tau}$,
is one-to-one and will be explicitly constructed below. The
constant $\varepsilon_0'$ here, just like $\varepsilon_0$ in (P5),
is assumed to be small enough, and moreover
\be
         \varepsilon_0' < C_0'\varepsilon_0
         \label{ee0}
\ee
with some constant $C_0'>0$.

We now derive elementary but important consequences of the above
assumptions. Since the density $\pi(y,v,\tau)$ vanishes for
$|v|<v_{\min}'$, so does the function $q(v,\tau;Y,W)$ defined by
(\ref{qp}). Moreover, for all $|W|<v_{\min}'$, the function
$q(v,\tau;Y,W)$ will be independent of $W$, and so we can write it
as $q(v,\tau;Y)$. Also, equation (\ref{quadraticint}) can be
simplified: the factor sgn$(v-W)$ can be replaced by sgn$\, v$.
Then, expanding the square in (\ref{quadraticint}) reduces it to a
quadratic equation for $W$:
\be
       Q_0W^2-2Q_1W+Q_2 = 0
         \label{quadratic}
\ee
where
\be
     Q_{0}=\int {\rm sgn}\, v\cdot q(v,\tau;Y)\, dv
        \label{Q0}
\ee
\be
     Q_{1}=\int v\, {\rm sgn}\, v\cdot q(v,\tau;Y)\, dv
        \label{Q1}
\ee
\be
     Q_{2}=\int v^2\, {\rm sgn}\, v\cdot q(v,\tau;Y)\, dv
        \label{Q2}
\ee
with $Y=Y(\tau)$. The integrals $Q_0,Q_1,Q_2$ have the following
physical meaning:
$$
        m Q_0=m_L-m_R
$$
$$
        m Q_1=p_L-p_R
$$
$$
        m Q_2=2(e_L-e_R)
$$
where $m_L,p_L,e_L$ represent the total mass, momentum and energy
of the incoming gas particles (per unit length) on the left hand
side of the piston, and  $m_R,p_R,e_R$ -- those on the right hand
side of it. The value $Q_2$ also represents the net pressure
exerted on the piston by the gas as if the piston did not move. Of
course, if $Q_2(\tau)=0$, then we must have $W(\tau)=0$, which
agrees with (\ref{quadratic}).

Next, under the above requirements on $\pi(y,v,\tau)$, the
function $q(v,\tau;Y)$ is, in a certain sense, nearly symmetric in
$v$ about $v=0$ (see \cite{CLS} for details). This fact implies
that $Q_0$ and $Q_2$ are small, more precisely
\be
       \max\{|Q_0|,|Q_2|\}\leq C'\varepsilon_0
            \label{Q0Q2}
\ee
where $C'>0$ is a constant depending on the parameters
$D_1',K_1'$, etc., but not on $\varepsilon_0$. At the same time,
the assumption (P3) guarantees that
\be
             Q_1\geq Q_{1,\min}>0
            \label{Q1min}
\ee
where $Q_{1,\min}$ is a constant depending on $\pi_{\min}',
v_1',v_2'$, etc., but  not on $\varepsilon_0$.

If $\varepsilon_0$ is small enough, there is a unique root of the
quadratic polynomial (\ref{quadratic}) on the interval
$(-v_{\min}',v_{\min}')$, which corresponds to the only solution
of (\ref{quadraticint}). Since this root is smaller, in absolute
value, than the other root of (\ref{quadratic}), it can be
expressed by
\be
         W(\tau)=\frac{Q_1-\sqrt{Q_1^2-Q_0Q_2}}{Q_0}
                   \label{Wroot}
\ee
where the sign before the radical is ``$-$'', not ``$+$''. Of
course, (\ref{Wroot}) applies whenever $Q_0\neq 0$, while for
$Q_0=0$ we simply have $W(\tau)=Q_2/2Q_1$.

Eqs.\ (\ref{Q0Q2})-(\ref{Wroot}) imply an upper bound on the
piston velocity: $|W(\tau)|\leq B'\varepsilon_0$ for some constant
$B'>0$ depending on $D_1',K_1'$, etc., but  not on
$\varepsilon_0$. A similar bound holds for the piston acceleration
$A(\tau)=\dot{W}(\tau)$, since
$$
       A(\tau)=\frac{(dQ_0/d\tau)W^2-2(dQ_1/d\tau)W
       +(dQ_2/d\tau)}{2(Q_1-Q_0W)}
$$
and $|dQ_i/d\tau|=|(dQ_i/dY)W|\leq\,$const$\cdot\varepsilon_0$,
see \cite{CLS} for more details.

Next we consider the evolution of a point $(y,v)$ in the domain $
G:=\{(y,v):\ 0\leq y\leq 1\}$ under the rules (H1)--(H3), i.e.\ as
it moves freely with constant velocity and collides elastically
with the walls and the piston. Denote by $(y_{\tau},v_{\tau})$ its
position and velocity at time $\tau\geq 0$. Then (H1) translates
into $\dot{y}_{\tau} =v_{\tau}$ and $\dot{v}_{\tau}=0$ whenever
$y_{\tau} \notin\{0,1,Y(\tau)\}$, (H2) becomes
$(y_{\tau+0},v_{\tau+0})= (y_{\tau-0},-v_{\tau-0})$ whenever
$y_{\tau-0} \in\{0,1\}$, and (H3) gives
\be
       (y_{\tau+0},v_{\tau+0})=(y_{\tau-0},2W(\tau)-v_{\tau-0})
          \label{tildeFpiston}
\ee
whenever $y_{\tau-0}=Y(\tau)$. Note that (\ref{tildeFpiston})
corresponds to a special case of the mechanical collision rules
(\ref{V'})--(\ref{v'}) with $\varepsilon=0$ (equivalently, $m=0$).
Hence the point $(y,v)$ moves in $G$ as if it was a gas particle
with zero mass.

The motion of points in $(y,v)$ is described by a one-parameter
family of transformations $F^{\tau}:\, G\to G$ defined by
$F^{\tau}(y_0,v_0)=(y_{\tau},v_{\tau})$ for $\tau>0$. We will also
write $F^{-\tau}(y_{\tau},v_{\tau})=(y_0,v_0)$. According to
(H1)--(H3), the density $\pi(y,v,\tau)$ satisfies a simple
equation
\be
       \pi(y_{\tau},v_{\tau},\tau)=
       \pi(F^{-\tau}(y_{\tau},v_{\tau}),0)
       =\pi_0(y_0,v_0)
         \label{pFp}
\ee
for all $\tau\geq 0$. Also, it is easy to see that for each
$\tau>0$ the map $F^{\tau}$ is one-to-one and preserves area, i.e.
${\rm det}\,|DF^{\tau}(y,v)|=1$.

Now, because of (P4), the initial density $\pi_0(y,v)$ can only be
positive in the region
$$
     G^+:=\{ (y,v):\  0\leq y\leq 1,\ v_{\min}\leq |v|\leq v_{\max}\}
$$
hence we will restrict ourselves to points $(y,v)\in G^+$ only. At
any time $\tau>0$, the images of those points will be confined to
the region $G^+(\tau):=F^{\tau}(G^+)$. In particular,
$\pi(y,v,\tau)=0$ for $(y,v)\notin G^+(\tau)$.

The map $R_{\tau}:(y,v)\mapsto (y_{\ast},v_{\ast})$ involved in
(\ref{Rtau}) and (\ref{symmetry}) can now be defined as
$R_{\tau}=F^{\tau}\circ R_0\circ F^{-\tau}$, where
$R_0(y,v)=(1-y,-v)$ is a simple reflection ``across the piston''
at time $\tau=0$.

We now make an important observation. If a fast point
$(y_{\tau},v_{\tau})$ collides with a slow piston, $|W(\tau)|\ll
|v_{\tau}|$, they cannot recollide too soon: the point must travel
to a wall, bounce off it, and then travel back to the piston
before it hits it again.

Therefore, as long as (P1)--(P4) hold, the collisions of each
moving point $(y_{\tau},v_{\tau})\in G^+(\tau)$ with the piston
occur at well separated time moments, which allows us to
effectively count them. For $(x,v)\in G^+$
$$
     N(y,v,\tau)=\#\{s\in (0,\tau):\ y_s=Y(s),\ v_s\neq W(s)\}
$$
is the number of collisions of the point $(y,v)$ with the piston
during the interval $(0,\tau)$. For each $\tau>0$, we partition
the region $G^+(\tau)$ into subregions
$$
     G_{n}^+(\tau):=\{ F^{\tau}(y,v):\,
     (y,v)\in  G^+\ \ \&\ \ N(y,v,\tau)=n\}
$$
so $G^+_{n}(\tau)$ is occupied by the points that at time $\tau$
have experienced exactly $n$ collisions with the piston during the
interval $(0,\tau)$.

Now, for each $n\geq 1$ we define $\tau_n>0$ to be the first time
when a point $(y_{\tau},v_{\tau})\in G^+(\tau)$ experiences its
$(n+1)$-st collision with the piston, i.e.
$$
   \tau_n=\sup\{\tau>0:\, G_{n+1}^+(\tau) = \emptyset\}
$$
In particular, $\tau_1>0$ is the earliest time when a point
$(y_{\tau},v_{\tau})\in G^+(\tau)$ experiences its first
recollision with the piston. Hence, no recollisions occur on the
interval $[0,\tau_1)$, and we call it the {\em zero-recollision
interval}. Similarly, on the interval $(\tau_1,\tau_2)$ no more
than one recollision with the piston is possible for any point,
and we call it the {\em one-recollision interval}.

The time moment $\tau_{\ast}$ mentioned in Theorem~\ref{tmmain} is
the earliest time when a point $(y_{\tau},v_{\tau})\in G^+(\tau)$
either experiences its third collision with the piston or has its
second collision with the piston given that the first one occurred
after $\tau_1$. Hence, $\tau_{\ast}\leq \tau_2$, and actually
$\tau_{\ast}$ is very close to $\tau_2$, see below.

The following theorem summarizes the properties of the solutions
of the hydrodynamical equations (H1)--(H4).

\begin{theorem}[\cite{CLS}]
Let $T>0$ be given. If the initial density $\pi_0(y,v)$ satisfies
{\rm (P1)--(P5)} with a sufficiently small $\varepsilon_0$, then
\begin{itemize}
\item[{\rm (a)}] the solution of our hydrodynamical equations {\rm
(H1)--(H4)} exists and is unique on the interval $(0,T)$;
\item[{\rm (b)}] the density $\pi(y,v,\tau)$ satisfies conditions
similar to {\rm (P1)--(P4)} for all $0<\tau<T$, it also satisfies
(\ref{Y05})--(\ref{ee0}); \item[{\rm (c)}] The piston velocity and
acceleration remain small, $|W(\tau)|=O(\varepsilon_0)$ and
$|A(\tau)|=O(\varepsilon_0)$; \item[{\rm (d)}] we have
$|\tau_k-k/v_{\max}|=O(\varepsilon_0)$ for all $1\leq k<
Tv_{\max}$, and if $Tv_{\max}>2$, then also
$|\tau_{\ast}-\tau_2|=O(\varepsilon_0)$.
\end{itemize}
\label{tmprop}
\end{theorem}

\begin{corollary}
If $\varepsilon_0=0$, so that the initial density $\pi_0(y,v)$ is
completely symmetric about the piston, the solution is trivial:
$Y(\tau)\equiv 0.5$ and $W(\tau)\equiv 0$ for all $\tau>0$.
\label{crprop}
\end{corollary}

Lastly, we demonstrate the reason for  our assumption that all the
discontinuity curves of the initial density $\pi_0(y,v)$ must have
positive slopes. It would be quite tempting to let $\pi_0(y,v)$
have more general discontinuity lines, e.g. allow it be smooth for
$v_{\min}<|v|<v_{\max}$ and abruptly drop to 0 at $v=v_{\min}$ and
$v=v_{\max}$. The following example shows why this is not
acceptable.

\begin{figure}[h]
\centering
\epsfig{figure=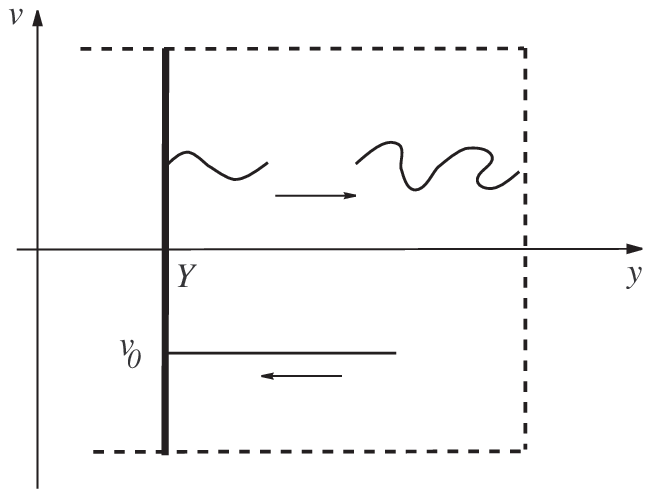}\caption{A horizontal discontinuity
line (bottom) comes off the piston as an oscillating curve (top).}
\end{figure}

\medskip\noindent{\bf Example}. Suppose the initial density
$\pi_0(y,v)$ has a horizontal discontinuity line $v=v_0$ (say,
$v_0=v_{\min}$ or $v_0=v_{\max}$). After one interaction with the
piston the image of this discontinuity line can oscillate up and
down, due to the fluctuations of the piston acceleration (Fig.~1).
As time goes on, this oscillating curve will ``travel'' to the
wall and come back to the piston, experiencing some distortions on
its way, caused by the differences in velocities of its points
(Fig.~1). When this curve comes back to the piston again, it may
well have ``turning points'' where its tangent line is vertical,
or even contain vertical segments of positive length. This
produces unwanted singularities or even discontinuities of the
piston velocity and acceleration. The same phenomena can also
occur when a discontinuity line of the initial density
$\pi_0(y,v)$ has a negative slope.

\section{Sketch of the argument}
\label{secS} \setcounter{equation}{0}

Our proof of Theorem~\ref{tmmain} is based on large deviation
estimates for the Poisson random variable:

\begin{lemma}[\cite{CLS}]
Let $X$ be a Poisson random variable with parameter $\lambda>0$.
For any $b>0$ there is a $c>0$ such that for all $0<B<b\sqrt{\lambda}$
we have
$$
    P(|X-\lambda|>B\sqrt{\lambda})\leq 2e^{-cB^2}
$$
\label{lmPoi}
\end{lemma}

This shows that the probabilities of large deviations rapidly
decay, as they do for the Gaussian distribution.

The principal step in our proof of Theorem~\ref{tmmain} is the
velocity decomposition scheme described next. Let $V_L(t,\omega)$
be the velocity of the piston at time $t\geq 0$ for a random
configuration of particles $\omega\in\Omega_L$. Let $\Delta t>0$
be a small time increment. Then the law of elastic collision
(\ref{V'}) implies
\be
    V_L(t+\Delta t,\omega)=(1-\varepsilon)^kV_L(t,\omega) +
    \varepsilon\sum_{j=1}^k(1-\varepsilon)^{k-j}\cdot v_j
       \label{main}
\ee
Here $k=k(t,\Delta t,\omega)$ is the number of particles colliding with the piston
during the time interval $(t,t+\Delta t)$, and $v_j$ are their
velocities numbered in the order in which the particles collide.

We rearrange the formula (\ref{main}) as follows:
\be
    V_L(t+\Delta t,\omega)=(1-\varepsilon k)V_L(t,\omega)
    + \varepsilon\sum_{j=1}^k v_j+\chi^{(1)}+\chi^{(2)}
          \label{main1}
\ee
where
$$
     \chi^{(1)}=V_L(t,\omega)[(1-\varepsilon)^k-1+\varepsilon k]
$$
and
$$
     \chi^{(2)}=\varepsilon\sum_{j=1}^kv_j[(1-\varepsilon)^{k-j}-1]
$$
Let us assume that the fluctuations of the velocity
$V_L(s,\omega)$ on the interval $(t,t+\Delta t)$, are bounded by
some quantity $\delta V$:
\be
       \sup_{s\in (t,t+\Delta t)} |V_L(s,\omega) - V_L(t,\omega)| \leq \delta V
        \label{deltaV}
\ee
Consider two regions on the $x,v$ plane:
\be
      D_1=\left \{(x,v):\  \frac{v-V_L(t,\omega)-({\rm sgn}\, v)\, \delta V}{x-X_L(t,\omega)}
         < -\frac{1}{\Delta t},\ \ v_{\min}<|v|<v_{\max} \right \}
               \label{D1}
\ee
and
\be
      D_2=\left \{(x,v):\  \frac{v-V_L(t,\omega)+({\rm sgn}\, v)\, \delta V}{x-X_L(t,\omega)}
         < -\frac{1}{\Delta t},\ \ v_{\min}<|v|<v_{\max} \right \}
               \label{D2}
\ee
Each of them is a union of two trapezoids $D_i=D_i^+\cup D_i^-$,
$i=1,2$, where $D_i^-$ denotes the upper and $D_i^+$ the lower
trapezoid, see Fig.~2.

\begin{figure}[h]
\centering
\epsfig{figure=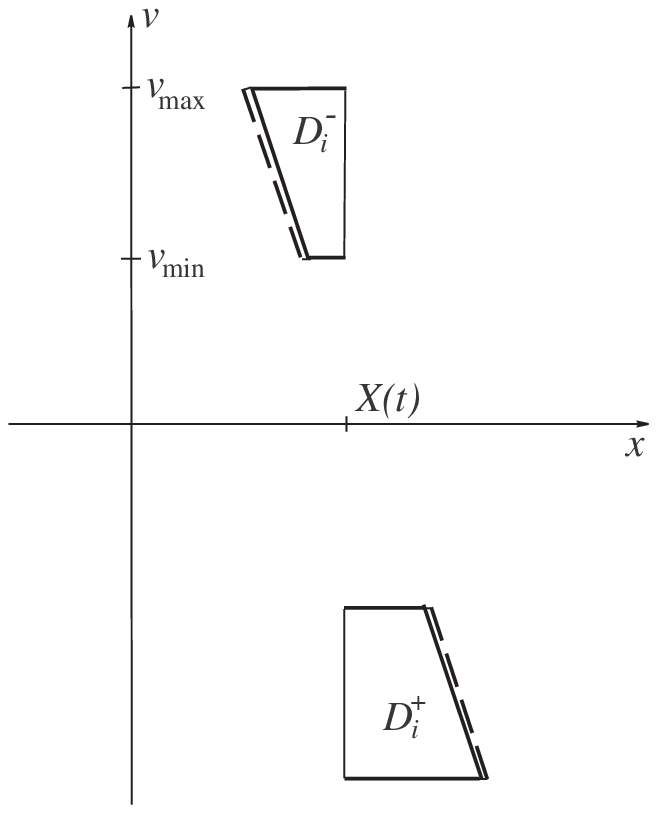} \caption{Region $D_1$ is bounded by
solid lines. Region $D_2$ is bounded by a dashed line.}
\end{figure}

Note that $D_1\subset D_2$. The bound (\ref{deltaV}) implies that
all the particles in the region $D_1$ necessarily collide with the
piston during the time interval $(t,t+\Delta t)$. Moreover, the
trajectory of every point $(x,v)\in D_1$ hits the piston within
time $\Delta t$. The bound (\ref{deltaV}) also implies that all
the particles actually colliding with the piston during the
interval $(t,t+\Delta t)$ are contained in $D_2$.

Let us denote by $k^{\pm}_r$ the number of particles in the regions
$D^{\pm}_r$ for $r=1,2$ at time $t$. We also denote
by $k^-$ the number of particles actually colliding with the piston ``on
the left'', and by $k^+$ that number ``on the right'' (of course,
$k^-+k^+=k$). Due to the above observations,
$k_1^{\pm}\leq k^{\pm}\leq k_2^{\pm}$.

Now, suppose that $t+\Delta t<\tau_1L$. Then we show that for
typical configurations $\omega$ the particles in each domain
$D_r$, $r=1,2$, have never collided with the piston before.
Therefore, their number, $k^{\pm}_r$, $r=1,2$, satisfies the laws
of Poisson distribution, in particular, the large deviation
estimate in Lemma~\ref{lmPoi} applies. This gives the bound (for
typical $\omega$)
\be
        \lambda_1^{\pm}-\Delta k_1^{\pm}
    \leq k^{\pm}\leq
    \lambda_2^{\pm}+\Delta k_2^{\pm}
       \label{lkl}
\ee
where
$$
        \lambda_r^{\pm}=E(k_r^{\pm})=
    L^2 \int_{F_L^{-t}(D^{\pm}_r)}p_L(x,v)\, dx\, dv
$$
and $F_L^{-t}$ corresponds to the action of $F^{-\tau}=F^{-t/L}$
in the original time-space coordinates $x,t$. The deviations
$\Delta k_r^{\pm}$ in (\ref{lkl}) can be adjusted by using
Lemma~\ref{lmPoi}. The difference $\lambda_2^{\pm}
-\lambda_1^{\pm}$ is estimated by
$$
        \lambda_2^{\pm}-\lambda_1^{\pm}=
    L^2 \int_{F_L^{-t}(D^{\pm}_2\setminus D^{\pm}_1)}
    p_L(x,v)\, dx\, dv \leq {\rm const}\cdot L^2\, \delta V\,\Delta t
$$
By putting all these estimates together we get tight bounds on
$k$ in (\ref{main1}). Similarly we get bounds on $\sum_{j=1}^k
v_j$ in (\ref{main1}). The following is the final result of this
analysis:
\be
        V_L(t+\Delta t,\omega)-V_L(t,\omega)=
        {\cal D}(t,\omega)\, \Delta t + \chi_3
           \label{VV}
\ee
Here
\be
       {\cal D}(t,\omega)=a\,
       [Q_0V_L^2(t,\omega)-2Q_1V_L(t,\omega)+Q_2]
          \label{calD}
\ee
and $Q_0,Q_1,Q_2$ are defined similarly to
(\ref{Q0})-(\ref{Q2}), in which $Y(\tau)$ must be replaced by the
actual piston position $X_L(t,\omega)/L$. The error term $\chi_3$
in (\ref{VV}) is bounded by
\be
        |\chi_3| \leq {\rm const}\cdot\frac{\ln L\,\sqrt{\Delta t}}{L}
       \label{chibound}
\ee
which corresponds to Brownian motion-type random fluctuations.

The term ${\cal D}(t,\omega)$ in (\ref{VV})
represents the main (``deterministic'') force acting on the
piston. The term $\chi_3$ describes random fluctuations
of that force. When the piston velocity stabilizes, then
the main force ${\cal D}$ should vanish, and an
``equilibrium'' velocity $\bar{V}_L(t,\omega)$ will
be established. The latter is the
root of the equation ${\cal D}(t,\omega)=0$, which is
\be
    \bar{V}_L(t,\omega)=\frac{Q_1-\sqrt{Q_1^2-Q_0Q_2}}{Q_0}
        \label{Vroot}
\ee
The reason why $V_L(t,\omega)$ converges to $\bar{V}_L$ is that
${\cal D}$ is almost proportional to $\bar{V}_L-V_L$, i.e.
$0<E_1<{\cal D}/(\bar{V}_L-V_L)<E_2<\infty$ for some constants
$E_1,E_2$. In fact, $\bar{V}_L(t,\omega)$ is a very slowly
changing function of $t$, whose derivative is small:
$|d\bar{V}_L(t,\omega)/dt|\leq {\rm const}\cdot
L^{-1}\varepsilon_0$. As a result, $V_L$ will always stay close to
$\bar{V}_L$, more precisely
\be
       |V_L(t,\omega)-\bar{V}_L(t,\omega)| < {\rm const}\cdot L^{-1}\,{\ln L}
             \label{VbarV}
\ee
on the entire zero-recollision interval $0<t<\tau_1L$.

Now, the piston coordinate $Y_L(\tau,\omega)=X_L(\tau L,\omega)/L$
is the solution of the differential equation
$$
        \dot{Y}_L=V_L=\bar{V}_L+\chi_4
$$
where $|\chi_4|<{\rm const}\cdot L^{-1}\,{\ln L}$ by
(\ref{VbarV}). On the other hand, the deterministic piston
coordinate $Y(\tau)$ is the solution of the equation $\dot{Y}=W$,
and both $\bar{V}$ and $W$ are given by the same radical
expression, cf.\ (\ref{Wroot}) and (\ref{Vroot}). Lastly, a simple
application of Gronwall's inequality completes the proof of
Theorem~\ref{tmmain} on the zero-recollision interval
$(0,\tau_1)$.

The proof on the one-recollision interval $(\tau_1,\tau_{\ast})$
goes along the same lines. One major difference is that the number
of particles $k^{\pm}_r$, $r=1,2$, in the domain $D^{\pm}_r$
constructed in the velocity decomposition scheme is no longer a
Poisson variable, so Lemma~\ref{lmPoi} does not apply directly.

To handle this new situation, we pull the domain $D^{\pm}_r$ back
in time, as we did before. But now that pullback involves one
interaction with the piston (corresponding to the first collision
of the particles in $D^{\pm}_r$ with the piston, which occurs
during the zero-recollision interval $0<t<\tau_1L$).

Since the piston position and velocity at the moment of that first
collision are random, the preimage of $D^{\pm}_r$ will be a random
domain. Its shape will depend on the piston velocity $V(t,\omega)$
during the zero-recollision interval $0<t<\tau_1L$. We observe
that the boundary of the preimage of $D^{\pm}_r$ is described by a
random, yet H\"older continuous function, and its H\"older
exponent is $0.5$ due to (\ref{chibound}). Then we pick a small
$d>0$ and construct a $d$-dense set in the space of all H\"older
continuous functions in the spirit of a work by Kolmogorov and
Tihomirov \cite{KT}. The elements of that $d$-dense set can be
used to construct a finite collection of (nonrandom) domains, so
that one of them will approximate the (random) preimage of our
$D^{\pm}_r$ (we need to select the small $d>0$ carefully to ensure
sufficient accuracy of the approximation). Now the number of
particles in our random domain (the preimage of $D^{\pm}_r$) can
be approximated by the number of particles in the corresponding
nonrandom domain. The latter has Poisson distribution, and finally
we can apply Lemma~\ref{lmPoi}. This trick gives necessary
estimates on $k^{\pm}_r$.

A full proof of Theorem~\ref{tmmain} is given in \cite{CLS}. At
present, we do not know if this theorem can be extended beyond the
critical time $\tau_{\ast}$, this is an open question. Some other
open problems are discussed in the next section.

\section{Discussion and open problems}
\label{secOPD} \setcounter{equation}{0}

1. The main goal of this work is to prove that under suitable
initial conditions random fluctuations in the motion of  a massive
piston are small and vanish in the thermodynamic limit. We are,
however, able to control those fluctuations effectively only as
long as the surrounding gas particles can be described by a
Poisson process, i.e.\ during the zero-recollision interval
$0<\tau<\tau_1$. In that case the random fluctuations are bounded
by const$\cdot L^{-1}\, \ln L$, see Remark~2 after
Theorem~\ref{tmmain}. Up to the logarithmic factor, this bound is
optimal, see \cite{CL} and earlier estimates by Holley \cite{H},
D\"urr et al.\ \cite{DGL}.

During the one-recollision interval $\tau_1<\tau<\tau_2$,  the
situation is different. The probability distribution of gas
particles that have experienced one collision with the piston is
no longer a Poisson process, it has intricate correlations. We are
only able to show that random fluctuations remain bounded by
$L^{-1/7}$, see again Remark~2. Perhaps, our bound is far from
optimal, but our numerical experiments reported in \cite{CL} show
that random fluctuations indeed grow during the one-recollision
interval.

We have tested numerically whether random fluctuations remained
small after more than one recollision, i.e.\ at times
$\tau>\tau_2$. We found that for some initial $\pi_0$ they
actually increased very rapidly, and we conjectured that the rate
of increase was exponential in $\tau$. We found, indeed, that at
times $\tau\sim \log L$ the fluctuations became large even on a
macroscopic scale, and then many unexpected phenomena occurred
\cite{CL}.

Interestingly, the exponential growth of random fluctuations seems
to be related to the instability of our hydrodynamical equations.
We found that small perturbations of the initial density $\pi_0$
can grow exponentially in $\tau$ under certain conditions,
matching the growth of random fluctuations of the piston motion in
the mechanical model. We refer the reader to \cite{CL} for further
discussion and to our work in progress \cite{CCLP}.

\medskip\noindent
2. It is clear that in our model recollisions of gas particles
with the piston have a very ``destructive'' effect on the dynamics
in the system. However, we need to distinguish between two types
of recollisions.

We say that a recollision of a gas particle with the piston is
{\em long} if the particle hits a wall $x=0$ or $x=L$ between the
two consecutive collisions with the piston. Otherwise a
recollision is said to be {\em short}. Long recollisions require
some time, as the particle has to travel all the way to a wall,
bounce off it, and then travel back to the piston before it hits
it again. Short recollisions can occur in rapid succession.

We have imposed the velocity cut-off (P4) in order to avoid any
recollisions for at least some initial period of time (which we
call the zero-recollision interval). More precisely, the upper
bound $v_{\max}$ guarantees the absence of long recollisions.
Without it, we would have to deal with arbitrarily fast particles
that dash between the piston and the wall very many times in any
interval $(0,\tau)$. On the other hand, the lower bound $v_{\min}$
was assumed to exclude short recollisions.

There are good reasons to believe, though, that short recollisions
may not be so destructive for the piston dynamics. Indeed, let a
particle experience two or more collisions with the piston in
rapid succession (i.e. without hitting a wall in between). This
can occur in two cases: (i) the particle's velocity is very close
to that of the piston, or (ii) the piston's velocity changes very
rapidly. The latter should be very unlikely, since the
deterministic acceleration of the piston is very small, cf.\
Theorem~\ref{tmprop}c. In case (i), the recollisions should have
very little effect on the velocity of the piston according to the
rule (\ref{V'}), so that they may be safely ignored, as it was
done already in earlier studies \cite{H,DGL}.

We therefore expect that our results can be extended to velocity
distributions without a cut-off from zero, i.e. allowing
$v_{\min}=0$.

\medskip\noindent
3. In our paper, $L$ plays a dual role: it parameterizes the mass
of the piston ($M\sim L^2$), and it represents the length of the
container ($0\leq x\leq L$). This duality comes from our
assumption that the container is a cube.

However, our model is essentially one-dimensional, and the mass of
the piston $M$ and the length of the interval $0\leq x\leq L$ can
be treated as two independent parameters. In particular, we can
assume that the container is infinitely long in the $x$ direction
(so, {\em that} $L$ is infinite), but the mass of the piston is
still finite and given by $M\sim L^2$. In this case there are no
recollisions with the piston, as long as its velocity remains
small. Hence, our zero-recollision interval is effectively
infinite. As a result, Theorem~\ref{tmmain} can be extended to
arbitrarily large times. Precisely, for any $T>0$ we can prove the
convergence in probability:
$$
   P\left (\sup_{0\leq\tau\leq T}
        |Y_L(\tau,\omega) - Y(\tau)| \leq C_T\ln L/L\right )\to 1
$$
and
$$
   P\left (\sup_{0\leq\tau\leq T}
        |W_L(\tau,\omega) - W(\tau)| \leq C_T\ln L/L\right )\to 1
$$
as $L\to\infty$, where $C_T>0$ is a constant and $Y(\tau)$ and
$W(\tau)=\dot{Y}(\tau)$ are the solutions of the hydrodynamical
equations described in Section~2.

\medskip\noindent
4. Along the same lines as above, we can assume that the container
is $d$-dimensional with $d\geq 3$. Then the mass of the piston
and the density of the particles are proportional to $L^{d-1}$
rather than $L^2$.

When $d$ is large, the gas particles are very dense on the $x,v$
plane. This leads to a much better control over fluctuations of
the particle distribution and the piston trajectory. As a result,
Theorem~\ref{tmmain} can be extended to the $k$-recollision
interval $(\tau_k,\tau_{k+1})$, where $k\geq 1$ depends on $d$. It
can be shown that for any $k\geq 1$ there is a $d_k\geq 3$ such
that for all $d\geq d_k$ the convergence (\ref{YY}) and (\ref{WW})
holds with $\tau_{\ast}=\tau_k$. Therefore, a higher dimensional
piston is more stable than a lower dimensional one.

It would be interesting to investigate other modifications of our
model that lead to more stable regimes. For example, let the
initial density $\pi_0(y,v)$ of the gas depend on the factor
$a=\varepsilon L^2$ in such a way that
$\pi_{0}(y,v)=a^{-1}\rho(y,v)$, where $\rho(y,v)$ is a fixed
function. Then the particle density grows as $a\to 0$. This is
another way to increase the density of the particles, but without
changing the dimension. One may expect a better control over
random fluctuations in this case, too.

\medskip\noindent
{\bf Acknowledgements}.  N.~Chernov was partially supported by NSF
grant DMS-9732728. J.~Lebowitz was partially supported by NSF
grant DMR-9813268 and by Air Force grant F49620-01-0154. Ya.~Sinai
was partially supported by NSF grant DMS-9706794. This work was
completed when N.~C. and J.~L. stayed at the Institute for
Advanced Study with partial support by NSF grant DMS-9729992.

\end{document}